\documentclass[twocolumn, twoside]{article} 

\usepackage[english]{babel}
\usepackage{graphicx} 

\usepackage{latexsym} 
\usepackage{amsfonts} 
\usepackage{amsmath} 
\usepackage{mathtools} 
\usepackage{cases}
\usepackage{breqn}

\usepackage{abstract}

\usepackage[weather]{ifsym}  
\usepackage[utf8]{inputenc} 
\usepackage[T1]{fontenc}

\usepackage{layout}

\usepackage{subfig} 
\usepackage{caption}

\usepackage{titlesec} 
\titleformat{\paragraph} 
    {\normalfont\bfseries}
    {}
    {0pt}
    {}

\newcommand{\defin}{\stackrel{\text{ {\scriptsize def} }}{=}}

\DeclareMathSymbol{\minus} {\mathord}{operators}{"2D}

\newcommand\ket[1]{\left|#1\right\rangle}
\newcommand\bra[1]{\left\langle#1\right|}
\newcommand\ketbra[2]{\ket{#1}\bra{#2}}
\newcommand\braket[2]{\left\langle #1 | #2 \right\rangle}

\usepackage{ntheorem}  

\newtheorem{theorem} {Theorem} [section]

\begin{document}

\hyphenation{}

\vspace{1.5 cm}
\title{\bf{Introduction to quantum information theory and outline of two applications to physics: \\the black hole information paradox and \\the renormalization group information flow
 }}
 
 \author{Fabio Grazioso\\Research associate - Science Education at Università degli Studi di Napoli Federico II, Italy,\\
Postdoctoral fellow at INRS-EMT, 1650 Boulevard Lionel-Boulet, Varennes, Québec J3X 1S2, Canada.}

\date{}

\twocolumn[
\maketitle
\begin{onecolabstract}
This review paper is intended for scholars with different backgrounds, possibly in only one of the subjects covered, and therefore little background knowledge is assumed. 

The first part is an introduction to classical and quantum information theory (CIT, QIT): basic definitions and tools of CIT are introduced, such as the information content of a random variable, the typical set, and some  principles of data compression. Some  concepts and  results of QIT are then introduced, such as the qubit, the pure and mixed states, the Holevo theorem, the no-cloning theorem, and the quantum complementarity.

In the second part, two applications of QIT to open problems in theoretical physics are discussed.

The black hole (BH) information paradox is related to the phenomenon of the Hawking radiation (HR). Considering a BH starting in a pure state, after its complete evaporation only the Hawking radiation will remain, which is shown to be in a mixed state. This either describes a non-unitary evolution of an isolated system, contradicting the evolution postulate of quantum mechanics and violating the no-cloning theorem, or it implies that the initial information content can escape the BH, therefore contradicting general relativity. The progress toward the solution of the paradox is discussed.

The renormalization group (RG) aims at the extraction of the macroscopic description of a physical system from its microscopic  description. This passage from microscopic to macroscopic can  be described in terms of several steps from one scale to another, and is therefore formalized as the action of a group. The c-theorem proves the existence, under certain conditions, of a function which is  monotonically decreasing  along the group transformations. This result suggests an interpretation of this function as entropy, and its use to study the \emph{information flow} along the RG transformations.
\end{onecolabstract}
]

\tableofcontents

\section{Classical information theory}

Classical information theory has been introduced by Claude Shannon in 1948 \cite{Shannon-48a, Shannon-48b}. In this seminal work he has devised a quantitative definition of information content, and then other formal definitions of relevant quantities, in order to allow for a quantitative treatment of those and other related subjects. In the same seminal work he also demonstrated some important theorems which hold for such quantities. In this first section we give a summary of the main concepts of the classical information theory introduced by Shannon.

\subsection{Information content}
\label{sec:info}
The first important contribution of Shannon has been to address the question: ``What is information?''.
More precisely, he was looking for a way to measure the amount of information \emph{contained} in a given physical system.
This is a rather elusive concept, and it can depend on things difficult to quantify, things such as the context, and the observer background knowledge.

\begin{figure}[!htbp]
	\begin{center}
	\includegraphics[width=\columnwidth]{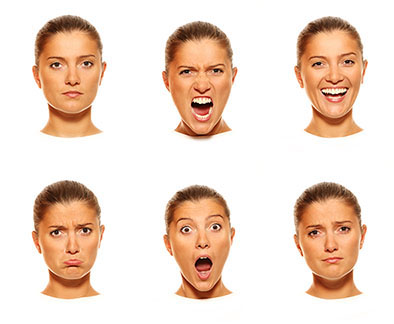}
	\caption{\label{fig:faces} Examples of facial expressions.}
	\end{center}
\end{figure}

To give an example, we can think at the amount of information contained in human facial expressions.
We know at an intuitive level that a big amount of information is contained in a single facial expression (see figure \ref{fig:faces}), since we sometimes take important decisions based on such informations. But at the same intuitive level we can appreciate how difficult is  to quantify this amount.
Moreover, the type of information in the example of the facial expressions refers to \emph{emotional states} or \emph{states of consciousness}, and therefore involve some degree of subjectivity in their definition (think e.g.\ at the famous painting ``Mona Lisa''  by Leonardo da Vinci, and its enigmatic facial expression, so difficult to define). As usual in science, Shannon has overcome this type of difficulty by first defining clearly the \emph{scope} of his definition.
His definition of ``content of information'' is indeed limited to systems that can be described by a \emph{random variable}.

Since we need a precise definition of random variable, following the notation of MacKay \cite{MacKay-03} we will use the concept of \emph{ensemble}, i.e.\ the collection of  three objects:
\begin{equation}
X \equiv \left(x, \mathcal{A}_{X}, \mathcal{P}_{X}\right)
\end{equation}
where $x$ represents the value of the random variable, $ \mathcal{A}_{X}$ is the set of the possible values it can assume, and $ \mathcal{P}_{X}$ is its \emph{probability distribution} of those values (i.e.\ the set of the probabilities of each possible value).

\subsubsection{Information content of a single outcome}
Based on this concept we then introduce the following definition for the \emph{amount of information} gained from the knowledge of a single outcome $x_{i}\in\mathcal{A}_{X}$ of the random variable $X$:
\begin{equation} \label{eq:info-single-outcome}
h(x_{i}) \equiv \frac{1}{\log 2} \log \frac1{p(x_{i})}
\end{equation}
where $p(x_{i})\in\mathcal{P}_{X}$ is the probability of the outcome $x_{i}$.
To give an intuition of this definition we can consider the example of the weather forecast. Let's simplify, and consider a situation where two only possible weather conditions are possible: \emph{sunny}  (\Sun) and \emph{rainy}  (\RainCloud). So, in our example the random variable is ``tomorrow's weather'',  the two possible values are $\mathcal{A}_{X} = \{\text{\Sun},\ \text{\RainCloud} \}$, and there will be a probability distribution $\mathcal{P}_{X} = \{p(\text{\Sun}),\  p(\text{\RainCloud})  \}$.

It is  worth noting that the definition of Shannon is totally independent from the actual value of the outcome, and only depends on its probability. It is in order to stress this concept that we have used the symbols $\{\text{\Sun},\ \text{\RainCloud} \}$ for the values of the outcome, that are not numerical, and do not appear at all in \eqref{eq:info-single-outcome}.
It is also worth to stress that  this definition of ``amount of information contained in a single outcome'' is a \emph{differential} definition: the difference between the amount of information we possess about the random variable, before and after we know the outcome.

We can illustrate this concept of ``differential definition'' using the weather variable: in a location where there is a very high probability of sunny weather, with the probability distribution $\mathcal{P}_{X} = \{p(\text{\Sun})=0.99,\ p(\text{\RainCloud}) = 0.01 \}$, if tomorrow  we see sunny weather, we will have learnt very little information. On the other hand, if tomorrow we find rainy weather, we will have gained a lot of useful information, with respect to today.

\subsubsection{Information content of a random variable}
Using the definition \eqref{eq:info-single-outcome} of the information content of a single outcome, we can define the information content of a whole random variable:
\begin{equation} \label{eq:Sh-entropy}
\begin{split}
H(X) &\equiv  \sum_{i} p(x_{i}) h(x_{i})\\
& = \frac{1}{\log 2} \sum_{i} p(x_{i}) \log \frac1{p(x_{i})}
\end{split}
\end{equation}
This definition can be seen as the \emph{average} of the information gained for each outcome expressed in \eqref{eq:info-single-outcome}, averaged over all the possible outcomes.

\begin{figure}[!htbp]
	\begin{center}
	\includegraphics[width=0.8\columnwidth]{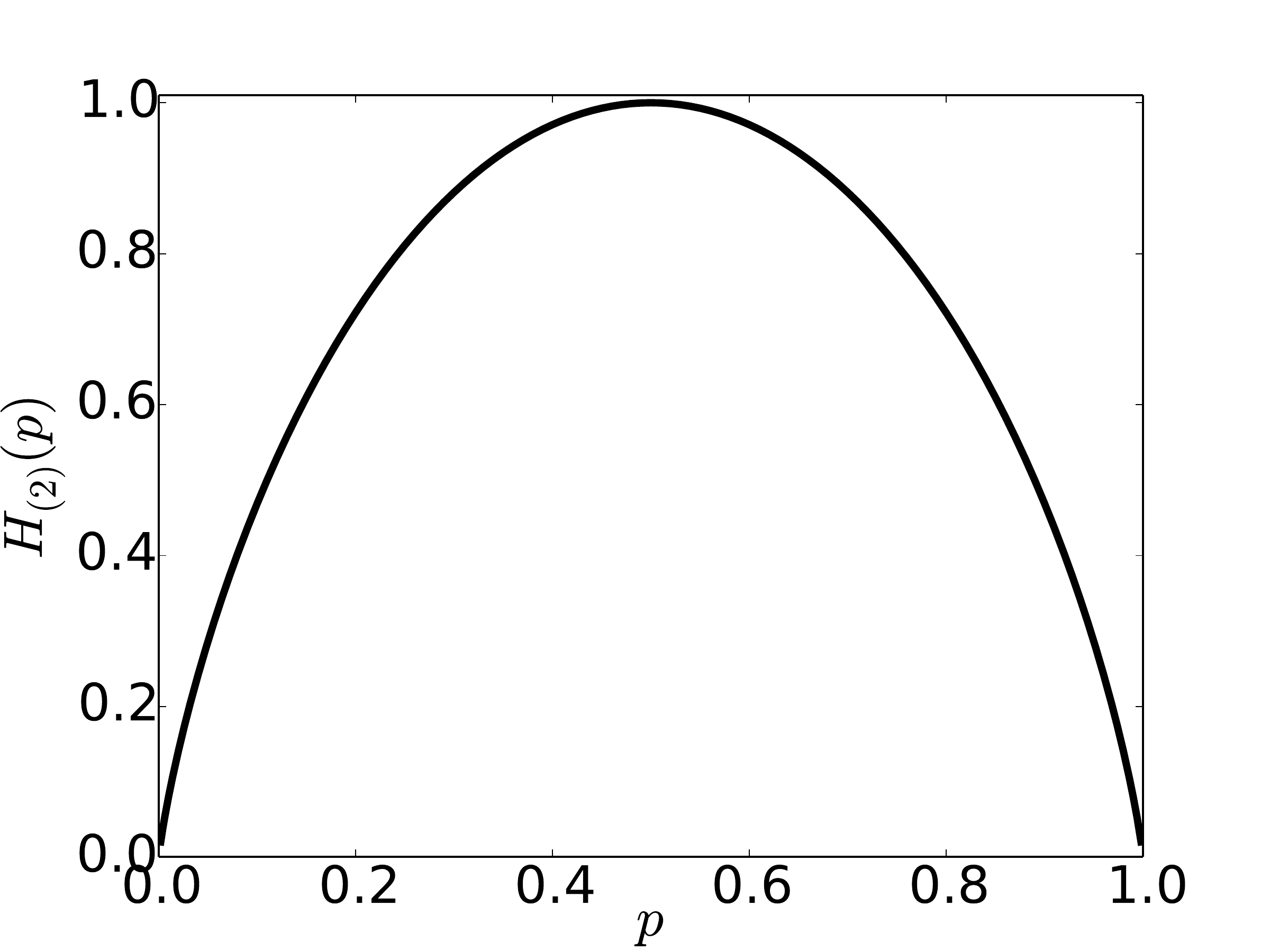}
	\caption{\label{fig:bin-entropy} Plot of the entropy of a binary variable (binary entropy) shown in \eqref{eq:bin-entropy}.}
	\end{center}
\end{figure}

This expression is formally equal (apart from constant factors) to the \emph{entropy} defined in thermodynamics, and Shannon proposed the same name in the context of information theory. This entropy is sometimes called ``Shannon entropy'', to distingush it from its quantum counterpart, discussed in the following.
In the case of a binary variable (i.e.\ variable with only two possible outcomes) we have:
\begin{subequations}
\begin{align}
&\mathcal{A}_{X} = \{0,1\}\\
&\mathcal{P}_{X} = \{p, (1-p)\},
\end{align}
\end{subequations}
and the entropy of a binary random variable gets the special name of \emph{binary entropy}:
\begin{equation} \label{eq:bin-entropy}
H_{(2)} = \frac{1}{\log 2} \ \left[p \log \frac1{p} + (1-p) \log \frac1{(1-p)}\right]
\end{equation}
A plot of the binary entropy as a function of $p$ is shown in  figure \ref{fig:bin-entropy}.

Again as for the information content of a single outcome, we can give some intuition for the definition of the entropy (i.e.\ information content) of a random variable using the example of the weather forecast. We can notice that in the case of a very biased probability distribution $\mathcal{P}_{X} = \{p(\text{\Sun})=0.01,\ p(\text{\RainCloud}) = 0.99 \}$, although the information content of the very unlikely outcome $h(\text{\Sun}) = \frac{1}{\log 2} \log \frac1{0.01}$ is very high, its weight (i.e.\ probability) in the average \eqref{eq:bin-entropy} is very small. So we have that \emph{the highest value for the binary entropy is for the uniform probability distribution} $\mathcal{P}_{X} = \{p(\text{\Sun})=0.5,\ p(\text{\RainCloud}) = 0.5 \}$, so that $p=1/2$ and all the outcomes are \emph{equiprobable}. It can be shown that this is true not only for the case of a binary variable, but for all the entropies of any random variable.
This also explains the constant factor $\frac1{\log 2}$ in the definitions of the entropies: it is a normalization factor, so that the maximum entropy is normalized to $1$. The factor $\frac1{\log 2}$ has also the advantage to make the definitions \eqref{eq:Sh-entropy} and \eqref{eq:bin-entropy} independent of the choice of the basis for the logarithms. Alternative and equivalent definitions are:
\begin{subequations}
\begin{align}
&H =  - \sum_{i} p(x_{i}) \log_{2} p(x_{i})\\
&H_{(2)} = - p \log_{2} p - (1-p) \log_{2} (1-p).
\end{align}
\end{subequations}
With this normalization is said that the entropy is measured in \emph{bits}, and the entropy of an unbiased binary variable is 1.
Sometimes another normalization is used, where the $\log_{2}$ is replaced by the natural logarithm $\ln = \log_{e}$; in this case it is said that the entropy is measured in \emph{nats}.

\subsubsection{comments}

We can find  an intuitive justification of the definition \eqref{eq:info-single-outcome} doing the following observations. First, the probability of two independent variables is the product of the probabilities of each outcome. On the other hand, for the definition \eqref{eq:info-single-outcome} of ``information from a single outcome'' it is reasonable that the information gained from two outcomes from two independent variables is the \emph{sum} of the information gained from each outcome. Thirdly, we have emphasized that the information content only depends on the probability. Given all this, when looking for an expression of the information content, the logarithm of the probability fits all the requirements. The last detail of using the logarithm of the \emph{inverse} of the probability is coming from the requirement that the entropy of a variable has to be maximal (and not minimal) in the case of uniform probability distribution (see figure \ref{fig:bin-entropy}).

\subsection{Other important definitions}

For the applications we want to introduce in the following sections, we need to define few more quantities. The definitions we need involve two random variables:
\begin{subequations}
\begin{align}
&\{X, \mathcal{A}_{X}, \mathcal{P}_{X} \}\\
&\{Y, \mathcal{A}_{Y}, \mathcal{P}_{Y} \}
\end{align}
\end{subequations}

\subsubsection{Joint entropy}
The joint probability $p(x,y)$ is defined as the probability that the variable $X$ has the outcome $x$ \emph{and} the variable $Y$ has the outcome $y$.
Based on this concept, it is easy to define the \emph{joint entropy} of two random variables as:
\begin{equation} \label{eq:joint-entropy}
H(X,Y_) \equiv \frac{1}{\log 2} \ \sum_{x,y} p(x,y) \log \frac{1}{p(x,y)}
\end{equation}
It is worth to recall from probability theory that the joint probability is the product of the probabilities in the case of \emph{independent random variables}. So in the case of independent variables the joint entropy is the sum of the entropies.

\begin{figure}[!htbp]
	\begin{center}
	\includegraphics[width=\columnwidth]{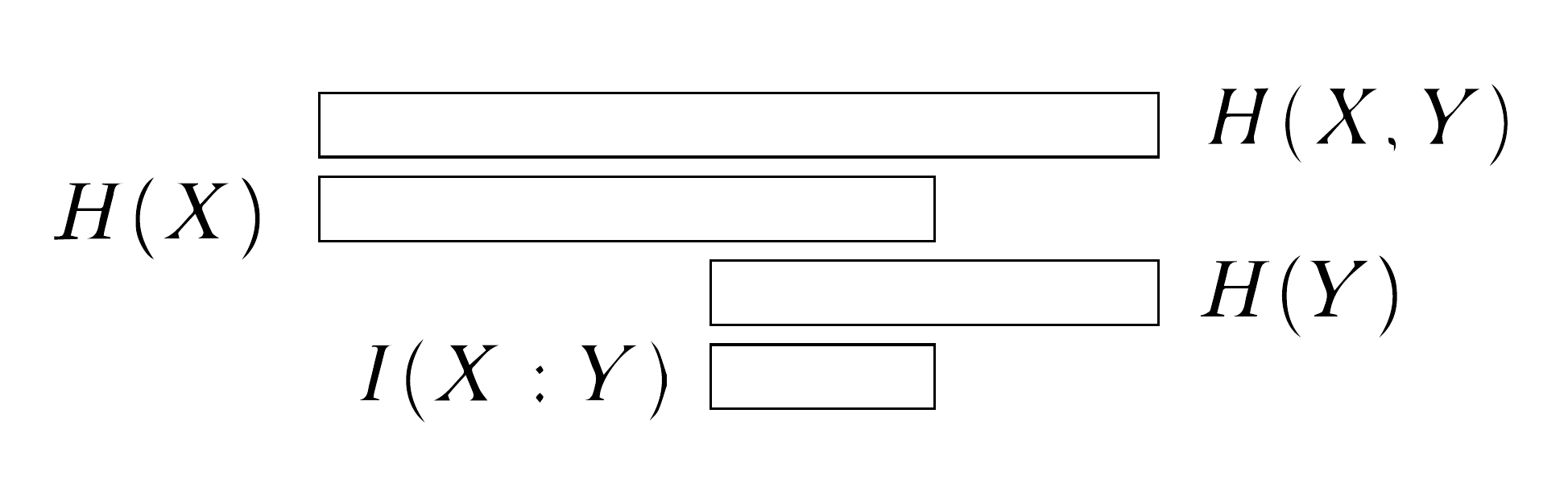}
	\caption{\label{fig:Venn} A graphical representation of the relationship between entropy, joint entropy and mutual information.}
	\end{center}
\end{figure}

Complementary to the concept of joint entropy is the definition of \emph{mutual information} of two random variables:
\begin{equation} \label{eq:mutu-info}
I(X:Y) \equiv H(X) + H(Y) - H(X,Y).
\end{equation}
We can use the intuition that \emph{mutual information is a measure of how much two random variables are not independent}.
It is also useful to rephrase this and think that \emph{mutual information is a measure of how much we know about a random variable $X$ if we know about random variable $Y$}.
It is frequently used a graphical representation to visualize the relationship between entropy, joint entropy and mutual information. Instead of the Venn diagrams \cite{Venn-880a, Venn-880b}, sometimes misleading, we prefer to use the alternative approach used e.g.\ by \cite{MacKay-03}, shown in figure \ref{fig:Venn}.

\subsection{Source coding theorem}
After having introduced some definitions, we here describe a theorem, called \emph{source coding theorem}. 

First, we have to introduce the notion of a source, described as a black box producing sequences of values. The way to model this is to consider those values as the outcomes of random variables. So we consider a sequence of $N$ random variables, and we assume the following hypotheses: that the variables are \emph{independent from each other},  that \emph{the set of possible values is identical} for alle the variables, and  finally that \emph{the probability distributions are identical}. This is usually summarized as the N variables being \emph{independent and identically distributed}, or i.i.d..

\subsubsection{Typical set}

Let's consider a sequence of N i.i.d.\ \emph{binary} variables. We can write the sequence of variables as $(X_{1}, X_{2}, \ldots, X_{N}) = X^{N}$, and a single outcome will be a sequence of values as $(x_{1}, x_{2}, \ldots, x_{N}) = x^{N}$, which in the case of a binary variable can be represented as a sequence of $N$ ones and zeroes. We can call  $\mathcal{A}_{X^{N}}$ the set of all the possible sequences, and we can write it down, (e.g.\ using the \emph{lexicographic} order) as follows:
\begin{equation}
\begin{split}
(0,0,0,&0,0,\ldots,0)\\
(1,0,0,&0,0,\ldots,0)\\
(0,1,0,&0,0,\ldots,0)\\
&\vdots\\
(1,1,1,&1,1,\ldots,1)\\
\end{split}
\end{equation}

Given all this, the source coding theorem proves the existence of a subset of  $\mathcal{A}_{X^{N}}$, called \emph{typical set}, with the property that "almost all" the information contained in the random variable is indeed contained in this subset. Moreover, the theorem proves that for a sequence of $N$ i.i.d.\ variables with entropy $H(X)$, the typical set has $2^{NH(X)}$ elements in it.
To be more precise, the theorem can be verbally stated as follows:
\begin{theorem} [Source coding theorem]
$N$ i.i.d.\ random variables each with entropy $H(X)$ can be compressed into more than $2^{NH(X)}$ bits with negligible risk of information loss, as $N \rightarrow \infty$; conversely if they are compressed into fewer than NH(X) bits it is ``virtually certain'' that some information will be lost.
\end{theorem}
It is of course possible to have a more precise statement, where instead of the ``almost all'' and ``virtually certain'' phrases, the proper mathematical expressions, with ``the epsilons and the deltas'' typical of the mathematical limits are used. For a proof of the theorem see e.g.\ \cite{MacKay-03,Cover-06}.

\subsubsection{Compression}

\begin{figure}[!htbp]
	\subfloat[\label{first-label}]
		{\includegraphics[width=\columnwidth]{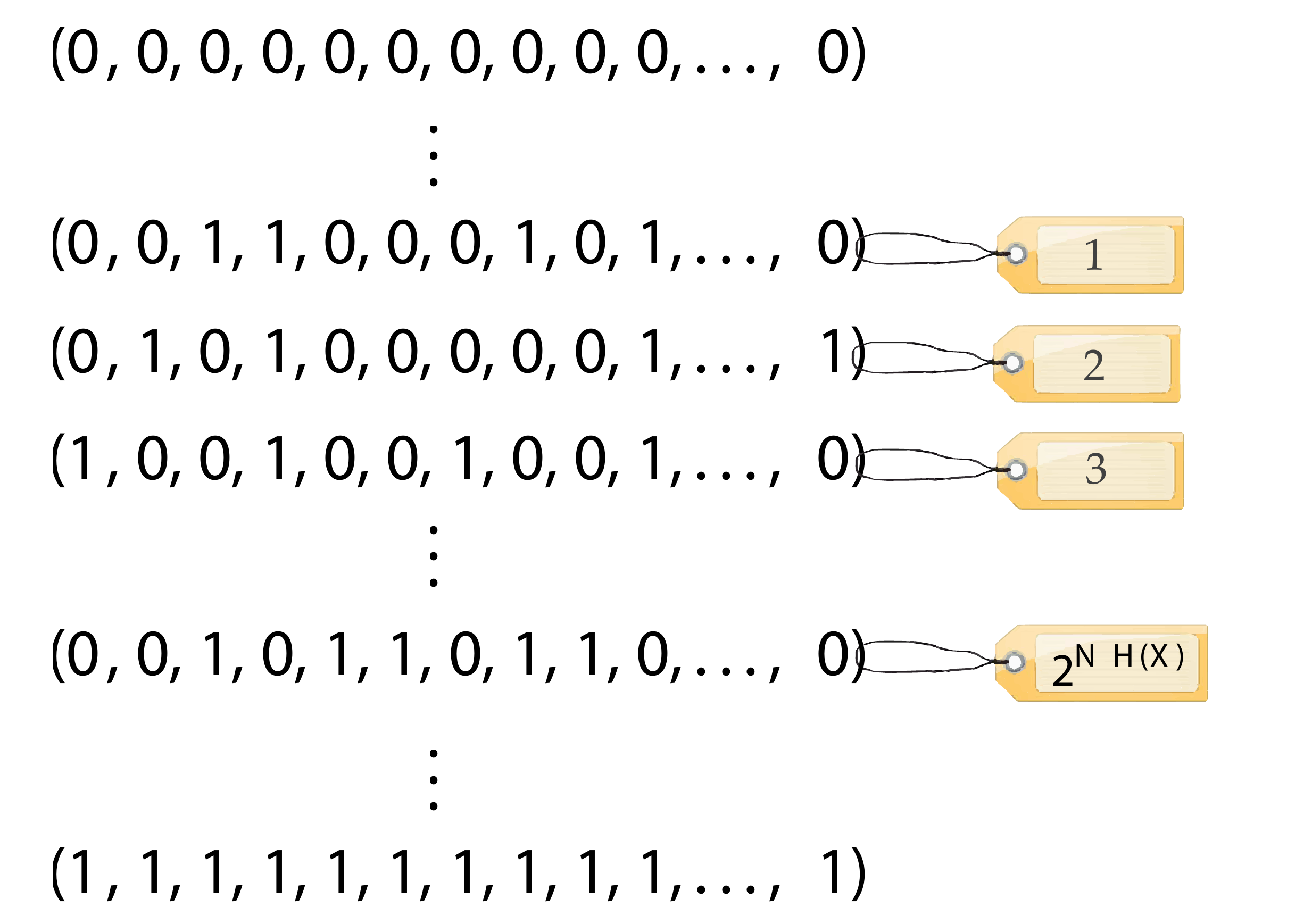}}
		\vspace{1cm}
		\\
	\subfloat[\label{second-label} ]
		{\includegraphics[width=\columnwidth]{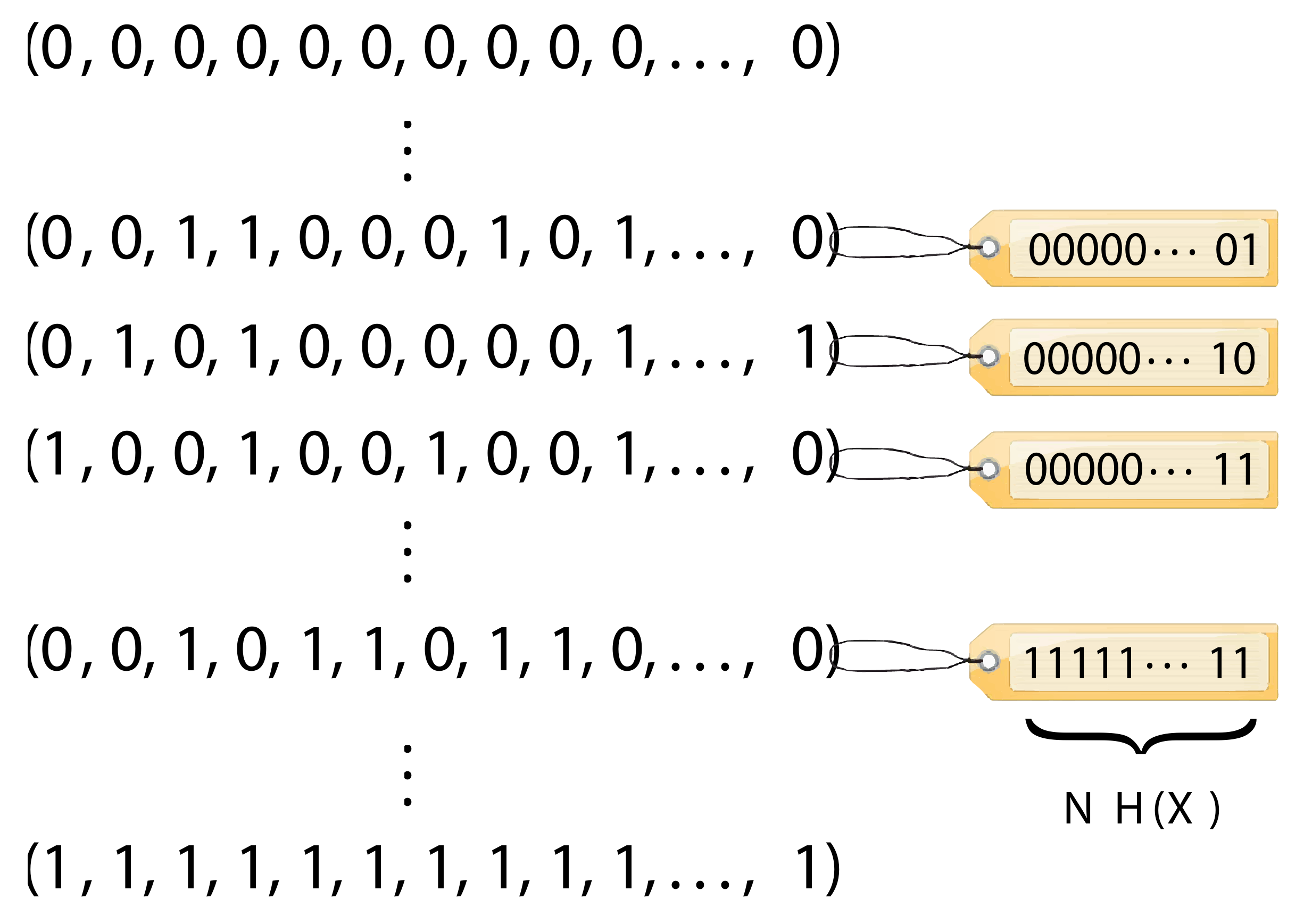}}
	\caption{\label{fig:typical-set} The typical set as a subset of all the possible sequences of N i.i.d. random variables outcomes. (a) The typical set elements can be labeled with a number between 1 and $2^{N H(X)}$. (b) This number can be written with $N H(X)$ binary simbols. }
\end{figure}

In figure \ref{fig:typical-set} we can see a graphical representation of the typical set, along with the idea that it is possible to label the elements of the typical set.
The fundamental idea of compression is that if we use only the $N H(X)$ symbols needed to  label  the elements of the typical set, instead of using the $N$ symbols of the full sequences, we have a negligible probability to loose information.

\section{Quantum Information Theory}

\begin{figure}[!htbp]
	\subfloat[\label{fig:qubit1} ]
		{\includegraphics[width=0.5\columnwidth]{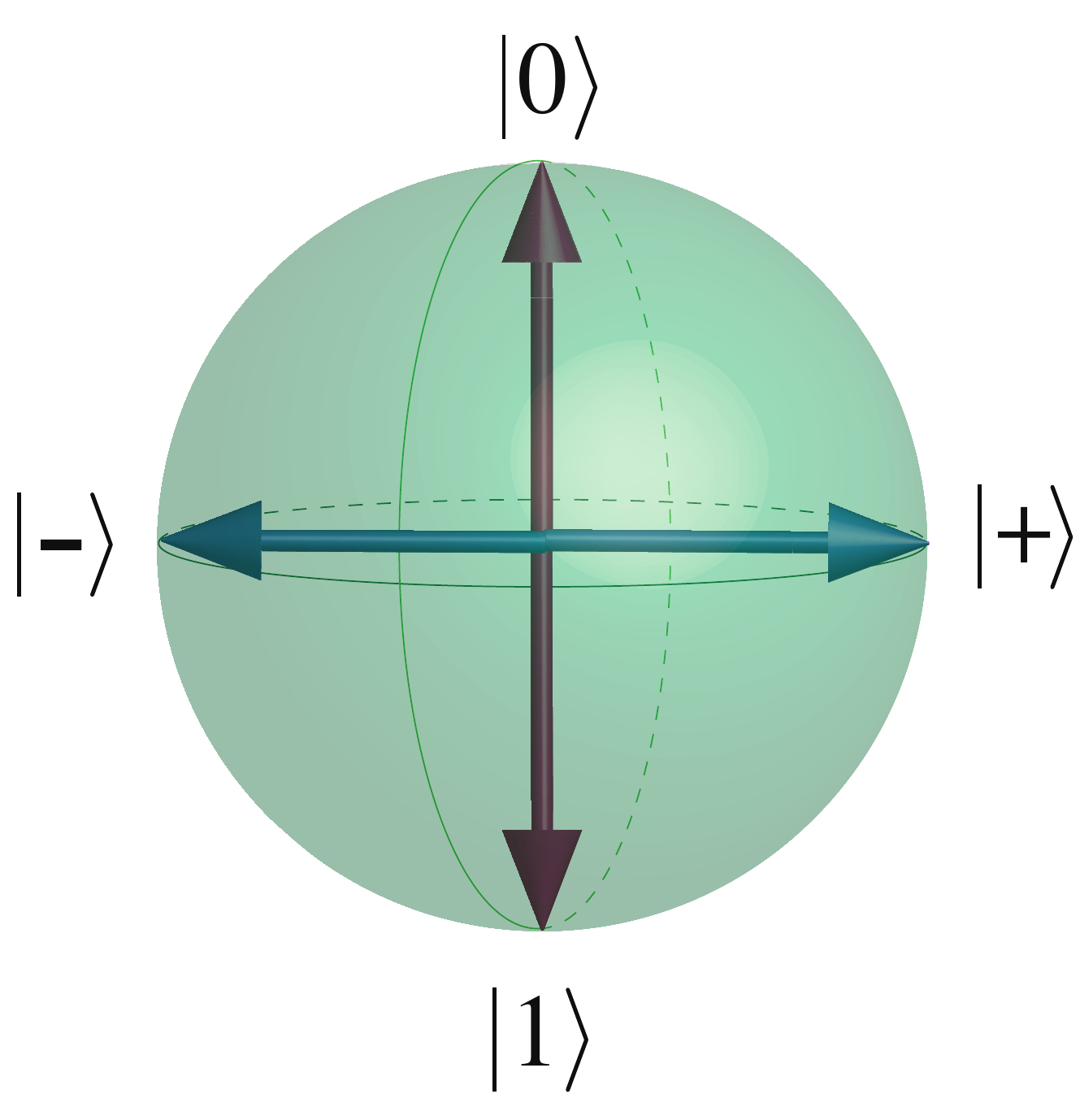}}
	\subfloat[\label{fig:qubit2} ]
		{\includegraphics[width=0.5\columnwidth]{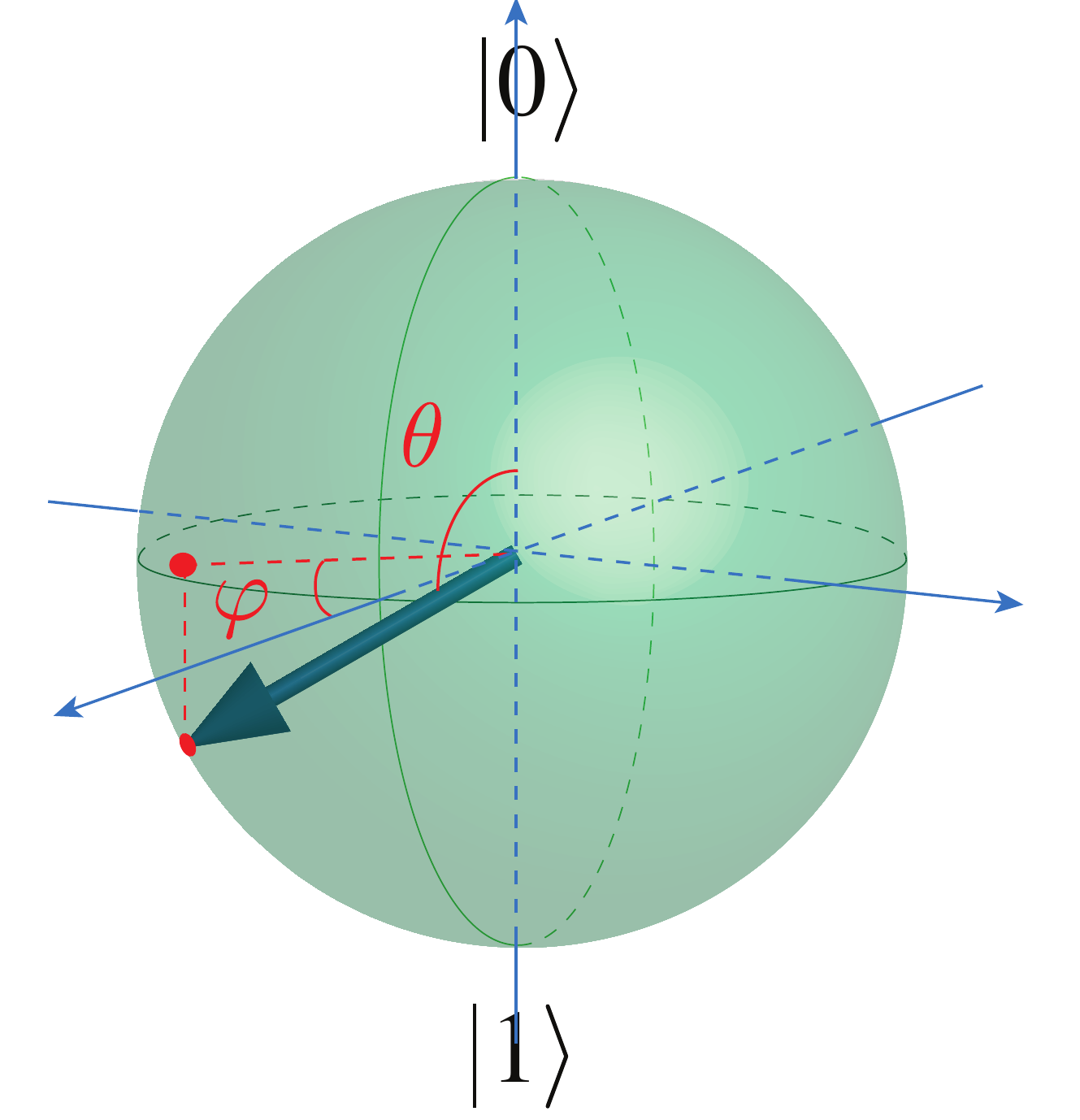}}
	\caption{\label{fig:qubit} The Block sphere is a two dimensional manifold, and is used to represent the two dimensional Hilbert space of the states of a qubit.}
\end{figure}

If the physical system used as support for the transmission and processing of information is a quantum system, classical information theory is no more  valid in all its parts, and a different theory has to be developed: quantum information theory (QIT).
As the classical random variable with two possible values (the bit) is the building block of CIT, the quantum random variable with its possible described by vectors of an Hilbert space of dimension two (the qubit) is the building block of QIT (see  figure \ref{fig:qubit}).
The experimental efforts to implement a qubit in a physical system have already a long history. Among the different approaches we can mention ion traps \cite{Cirac-95, Steane-00}, quantum dots \cite{Loss-98, Huibers-98}, nuclear spins, accessed via nuclear magnetic resonance \cite{DiVincenzo-95, Ernst-87}, colour defects in crystals \cite{Wrachtrup-01, Grazioso-13a} and superconductive structures \cite{Mooij-99, Yamamoto-03}.
In this section we will review the usual axiomatic introduction of quantum mechanics (QM) and the formal tools  which are necessary to describe the applications of QIT presented in the following. Among the many references for the axiomatic introduction to quantum mechanics, and the statements of its postulates, we refer mostly to \cite{Cohen1}.

\subsection{Mixed states and density operator formalism}

The state of a quantum system is represented by an element of an Hilbert space $\mathcal{H}$, of modulus one, which in the Dirac notation can be represented by a ``ket'' $\ket{\psi} \in \mathcal{H}$. In the case of a qubit (i.e.\ two-dimensional system) the basis can be represented as $\{\ket{0}, \ket{1}\}$ (called \emph{computational basis}), and the generic state of the qubit will be $\ket{\psi} = \alpha \ket{0} + \beta \ket{1}$, where $\alpha, \beta \in \mathbb{C}$, and the link to the angles shown in figure \ref{fig:qubit2} is  $\ket{\psi} = \cos \frac{\theta}{2} \ket{0} + e^{i\varphi} \sin \frac{\theta}{2} \ket{1}$.

In analogy to the concept of random variable introduced above, we need a formal tool to describe a situation where the state of the quantum system is unknown, and it is  only know the set of possible states, with their probability distribution.
If a system is in such conditions, it is said to be in a \emph{mixed state}, and the tool to describe mathematically a mixed state is the \emph{density operator}.

\subsubsection{Density operator of a pure state}

To introduce the density operator, let's first recall some details on linear algebra. The scalar product in the Dirac notation is written as $\langle\phi|\psi\rangle$; if we choose a basis $\{ \ket{1}, \ket{2}, \ldots, \ket{n}, \ldots\}$ of the Hilbert space, it is possible to compute the components $\langle i| \psi\rangle = \psi_{i}$ and $\langle \phi| i\rangle = \phi_{i}^{*}$ of the vectors and co-vectors, so to write them as  one-column and one-row matrices respectively. In this notation, the scalar product can be seen as a dot product between matrices:
\begin{subequations}
\begin{align}
\braket{\phi}{\psi}  &= (\phi_{1}, \phi_{2}, \ldots, \phi_{n}, \ldots)\left( \begin{array}{c} \psi_{1}\\ \psi_{2}\\ \vdots\\ \psi_{n} \end{array} \right)\\
&=\sum_{i} \phi_{i}^{*}\psi_{i}.
\end{align}
\end{subequations}

But if we invert the order, and write
\begin{subequations}
\begin{align}
\ketbra{\psi}{\phi} & = \left( \begin{array}{c} \psi_{1}\\ \psi_{2}\\ \vdots\\ \psi_{n} \end{array} \right)(\phi_{1}, \phi_{2}, \ldots, \phi_{n}, \ldots)\\
&= \left(
\begin{array}{cccc}
  \psi_{1}   \phi_{1}^{*} &  \psi_{1}   \phi_{2}^{*}   & \cdots &   \psi_{1}   \phi_{N}^{*}    \\
  \psi_{2}   \phi_{1}^{*} &   \psi_{2}   \phi_{2}^{*}   & \cdots &   \psi_{2}   \phi_{N}^{*}      \\
\cdots &  \cdots  & \cdots &   \cdots  \\
   \psi_{N}   \phi_{1}^{*} &   \psi_{N}   \phi_{2}^{*}   & \cdots &   \psi_{N}   \phi_{N}^{*}   
\end{array}
\right)
\end{align}
\end{subequations}
we have a matrix, which can be interpreted as the representation, in the chosen basis, of an \emph{operator defined on the same Hilbert space}. 

This was written for two different states $\ket\psi$ and $\ket\phi$. But using this type of product we can associate to any single vector of the Hilbert space an operator:
\begin{subequations}
\begin{align}
\ket{\psi} \ &\leftrightarrow \nonumber\\
&\leftrightarrow \  \ketbra{\psi}{\psi}  = \left( \begin{array}{c} \psi_{1}\\ \psi_{2}\\ \vdots\\ \psi_{n} \end{array} \right)(\psi_{1}, \psi_{2}, \ldots, \psi_{n}, \ldots)\\
&= \left(
\begin{array}{cccc}
  \psi_{1}   \psi_{1}^{*} &  \psi_{1}   \psi_{2}^{*}   & \cdots &   \psi_{1}   \psi_{N}^{*}    \\
  \psi_{2}   \psi_{1}^{*} &   \psi_{2}   \psi_{2}^{*}   & \cdots &   \psi_{2}   \psi_{N}^{*}      \\
\cdots &  \cdots  & \cdots &   \cdots  \\
   \psi_{N}   \psi_{1}^{*} &   \psi_{N}   \psi_{2}^{*}   & \cdots &   \psi_{N}   \psi_{N}^{*}   
\end{array}
\right)\\
& \defin \hat{\rho}_{\psi}\ .
\end{align}
\end{subequations}

\subsubsection{Density operator of a mixed state}
\label{sec:density-mixed}

When a state of a quantum system can be represented as a vector of an Hilbert space (i.e.\ a ket in Dirac notation), it is said to be in a \emph{pure state}. But if we want to represent the quantum analog of a random variable, we have to use the concept of mixed state introduced above, where we don't know the state of the system, but only  a set of possible states, and their respective probabilities. A mixed state for which all its possible states are equiprobable is said a \emph{maximally mixed state}.
It is interesting to point out that whether the system is in a pure or a mixed state depends on both the system \emph{and} the observer, because the knowledge about the system depends also on the observer, and not only on the system itself. 
The density operators formalism is able to effectively represent this type of states. 

Indeed, if the possible states of the system are $\left\{ \ket{\alpha_1}, \ket{\alpha_2}, \ldots, \ket{\alpha_N} \right\}$, with probabilities $\left\{p_1, p_2, \ldots, p_N \right\}$, then the  mixed state can be represented as:
\begin{equation}\label{eq:mixedstate}
\sum_{i=1} p_i \ketbra{\alpha_i}{\alpha_i}.
\end{equation} 
This can be seen as a linear combination of the density operators associated to the pure states, where the coefficients are the probabilities. 

This is an abstract representation of the density operators; if we fix a basis in the Hilbert space, we can write a density operator as a matrix, that will be called \emph{density matix}. 
A special and not uncommon case is when the set of possible states of a mixed state is an orthonormal basis for the Hilbert space. If we write this orthonormal basis as $\{\ket{1}, \ket{2}, \ldots, \ket{n}, \ldots \}$, and then represent the density matrix associated to a \emph{pure state} in this basis, the matrix elements will be all zero, apart from one single element on the diagonal equal to one, in the position corresponding to the position of the pure state in the basis:
\begin{align}
\hat{\rho}_n &= \ketbra{n}{n} = 
\left(
\begin{array}{c}
  0 \\
  \vdots \\
  n \\
  \vdots
\end{array} 
\right)
(0,\ldots, n, \ldots) \nonumber \\ 
&=
 \left(
\begin{array}{cccc}
  0&0&0&0 \\
  0&\ddots&0&0 \\
  0&0&1&0 \\
  0&0&0&\ddots
\end{array} \right).
\end{align}
If we then consider a  mixed state such that the possible states are all the elements of the basis:
\begin{equation}\label{eq:special-mixedstate}
\sum_{i=1} p_i \ketbra{i}{i}
\end{equation} 
its density matrix, represented in this same basis  will be diagonal,  with the probabilities as diagonal elements:
\begin{equation} \label{eq:diag-density}
\hat{\rho} =
 \left(
\begin{array}{cccc}
  p_1&0&0&0 \\
  0&\ddots&0&0 \\
  0&0&p_n&0 \\
  0&0&0&\ddots
\end{array} \right).
\end{equation}
If represented in this basis, non-zero off-diagonal elements indicate that some of the possible states are quantum superpositions of basis states.
From the normalization property of the probability distribution it is then easy to see that:
\begin{equation}
\text{Tr}(\hat{\rho}) = \sum_i p_i = 1,
\end{equation}
where $Tr(\hat{\rho})$ indicates the trace, defined as the sum of the diagonal elements. Since the trace is preserved under change of reference, we can conclude that $Tr(\hat{\rho})=1$ is a property of any density matrix.
Another property of any density matrix is that the eigenvalues are non-negative. This can be proven rigorously, and can be easily seen in the case of a diagonal density matrix \eqref{eq:diag-density}, where the eigenvalues have the meaning of probabilities.

\subsection{Quantum measurement and quantum complementarity}

Continuing with the  axiomatic introduction of quantum mechanics, after the concept of mixed states, and the density operators formalism to describe them, we  now describe  the measurement of the state of a quantum system. 

In the following subsections we will give two possible formalizations of the measurement process, namely the \emph{projective measurement}, and the \emph{POVM}.  Finally, we will see the concept of \emph{complementarity}.

\subsubsection{Projective measurement}

A first way to formalize the measurement process is the \emph{projective measurement} or \emph{von Neumann measurement} (see \cite{Cohen1, Nielsen-00}). In this description we associate to the measurement an hermitian operator $\hat{M}$, and its decomposition over the projectors on its eigenspaces:
\begin{equation} \label{eq:meas-op}
\hat{M} = \sum_{m} m \hat{P}_{m}
\end{equation}
where $\{m\}$, the eigenvalues of $\hat{M}$, are the possible outcomes of the measurement, and the $\{\hat{P}_{m}\}$ operators are projectors, i.e. satisfy the following properties:
\begin{subequations} \label{eq:proj-prop}
\begin{align}
&\forall m, \hat{P}_{m} \text{ is hermitian}\\
&\forall m, m', \hat{P}_{m} \hat{P}_{m'} = \delta_{m, m'} \hat{P}_{m}.
\end{align}
\end{subequations}
The probability that the outcome of the measurement is $m$ when the system is in the state $\ket\psi$ is:
\begin{equation} \label{eq:prob-project}
p_{\psi}(m) = \bra\psi \hat{P}_{m} \ket\psi;
\end{equation}
and soon after such measurement the state of the system is:
\begin{equation} \label{eq:state-after}
\frac{\hat{P}_{m} \ket\psi}{\sqrt{p_{\psi}(m)}}.
\end{equation}
From the requirement that the sum of all the probabilities \eqref{eq:prob-project} is equal to $1$ we have the property of \emph{completeness} for the set of projectors:
\begin{equation} \label{eq:completeness}
\sum_{m} \hat{P}_{m} = \mathbb{I}.
\end{equation}
The expectation value of the measurement $\hat{M}$ if the system is in the state $\ket\psi$ is:
\begin{equation}
\begin{aligned}
E_{\psi}(\hat{M}) & = \sum_{m} m\ p_{\psi}(m)\\
& = \sum_{m} m \bra{\psi}P_{m}\ket{\psi}\\
& =  \bra{\psi}\left(\sum_{m} m P_{m}\right) \ket{\psi}\\
& = \bra{\psi} \hat{M} \ket{\psi}\\
& = \langle \hat{M} \rangle_{\psi}.
\end{aligned}
\end{equation}
and the standard deviation is:
\begin{equation}
\begin{aligned}
\Delta(\hat{M}) &= \sqrt{\langle(\hat{M}-\langle \hat{M}\rangle_{\psi})^{2}\rangle_{\psi}}\\
&= \sqrt{\langle \hat{M}^{2}\rangle_{\psi} - \langle \hat{M}\rangle^{2}_{\psi}} 
\end{aligned}
\end{equation}
where we have used the compact notation $\bra\psi \cdot \ket\psi = \langle \cdot \rangle_{\psi}$.
Sometimes it is useful to write the projectors as:
\begin{equation} \label{eq:krauss}
\hat{P}_{m} = \hat{M}^{\dag}_{m} \hat{M}_{m}
\end{equation}
where $\hat{M}_{m}$ are called \emph{Krauss operators}. The  equations \eqref{eq:meas-op}-\eqref{eq:completeness} can be rewritten in terms of the Krauss operators using \eqref{eq:krauss}.

\subsubsection{POVMs} \label{sec:POVM}
It is possible to generalize  the projective measurement and define the POVM (positive operator-valued measurement \cite{Nielsen-00}), where some of the hypotheses of the projective measurement are relaxed. In particular, we consider the collection of operators that represent the measurement:
\begin{equation} \label{eq:POVM-ops}
\{\hat{E}_{m}\}
\end{equation}
and relax the hypothesis that those operators  are projectors.
Similarly to the projective measurement, the probability that the outcome is $m$ if the system is in $\ket\psi$ is:
\begin{equation}
p_{\psi}(m) = \bra{\psi}\hat{E}_{m}\ket{\psi}.
\end{equation}
Also for the POVM we have the  property of completeness: $\sum_m \hat{E}_m = \mathbb{\hat{I}}$, but as  a consequence of the \eqref{eq:POVM-ops} not being projectors, is that in general we can not write them in terms of the Krauss operators, as in \eqref{eq:krauss}, and therefore for the POVM measurement it is not defined the state of the system \emph{after the measurement}.

A common situation with POVM measurement is when we have a quantum system in a mixed state, where the set of possible states are represented by some vectors of the Hilbert space  $\{\ket{\psi_m}\}$, not necessarily orthogonal to each other, and we want a measurement in order \emph{to know in which of the states of the set the system is}. This POVM is represented by the  set of operators: 
\begin{equation} \label{eq:POVM-projectors}
\{\hat{E}_m = \ketbra{\psi_m}{\psi_m}\}.
\end{equation}
These last operators are indeed projectors; however, since the $\{\ket{\psi_m}\}$ are not necessarily orthogonal, this POVM \emph{is not} in general a projective measurement. In this type of POVM, since the set of states does not necessarily form a basis of the Hilbert space, the completeness property has in general to be guaranteed with suitable normalization coefficients.

\subsubsection{Quantum complementarity}
\label{sec:complementarity}

\begin{figure}[!htbp]
\subfloat[\label{fig:proiez1} ]
{\includegraphics[width=0.5 \columnwidth]{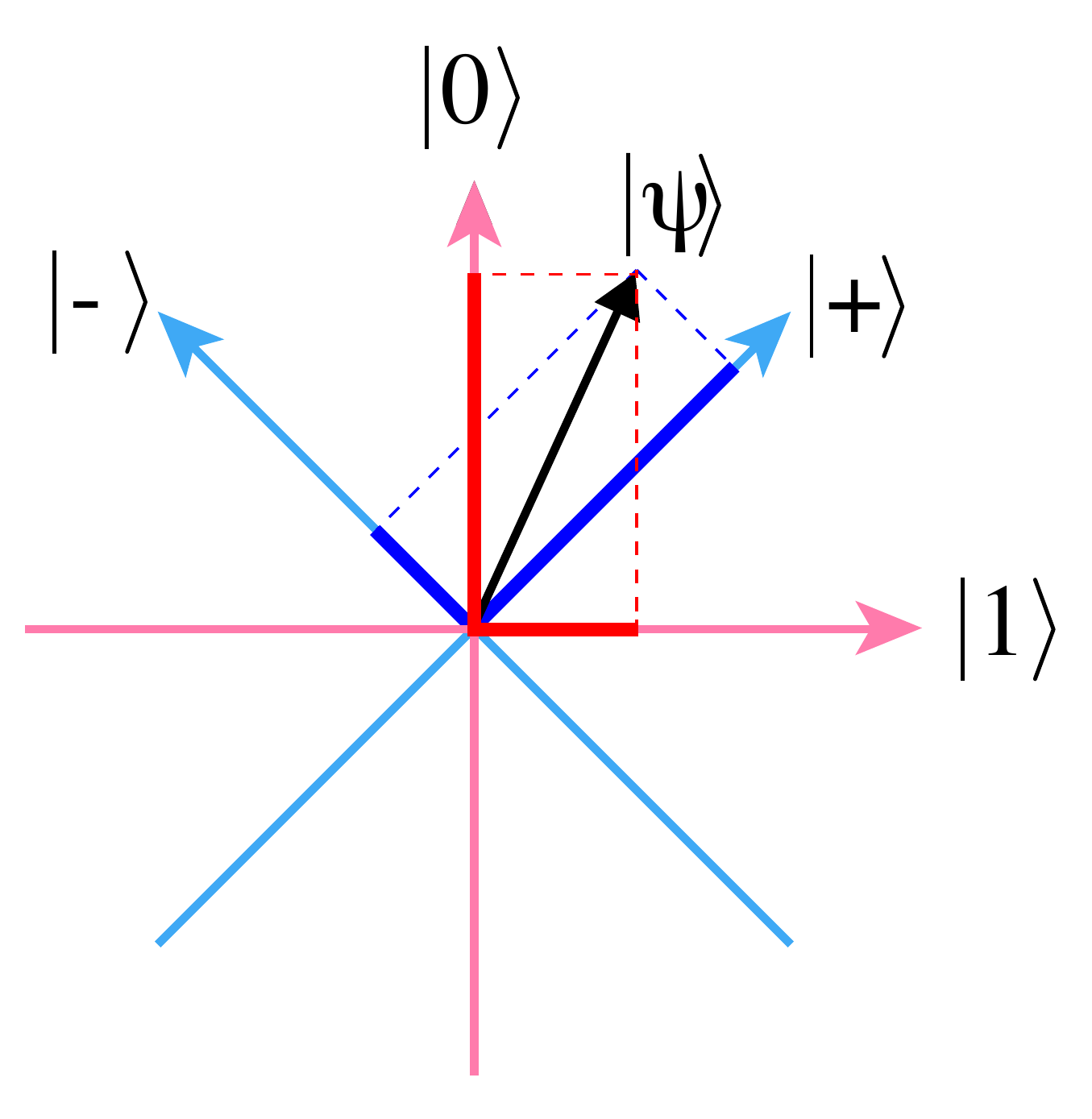}}
\subfloat[\label{fig:proiez2} ]
{\includegraphics[width=0.5 \columnwidth]{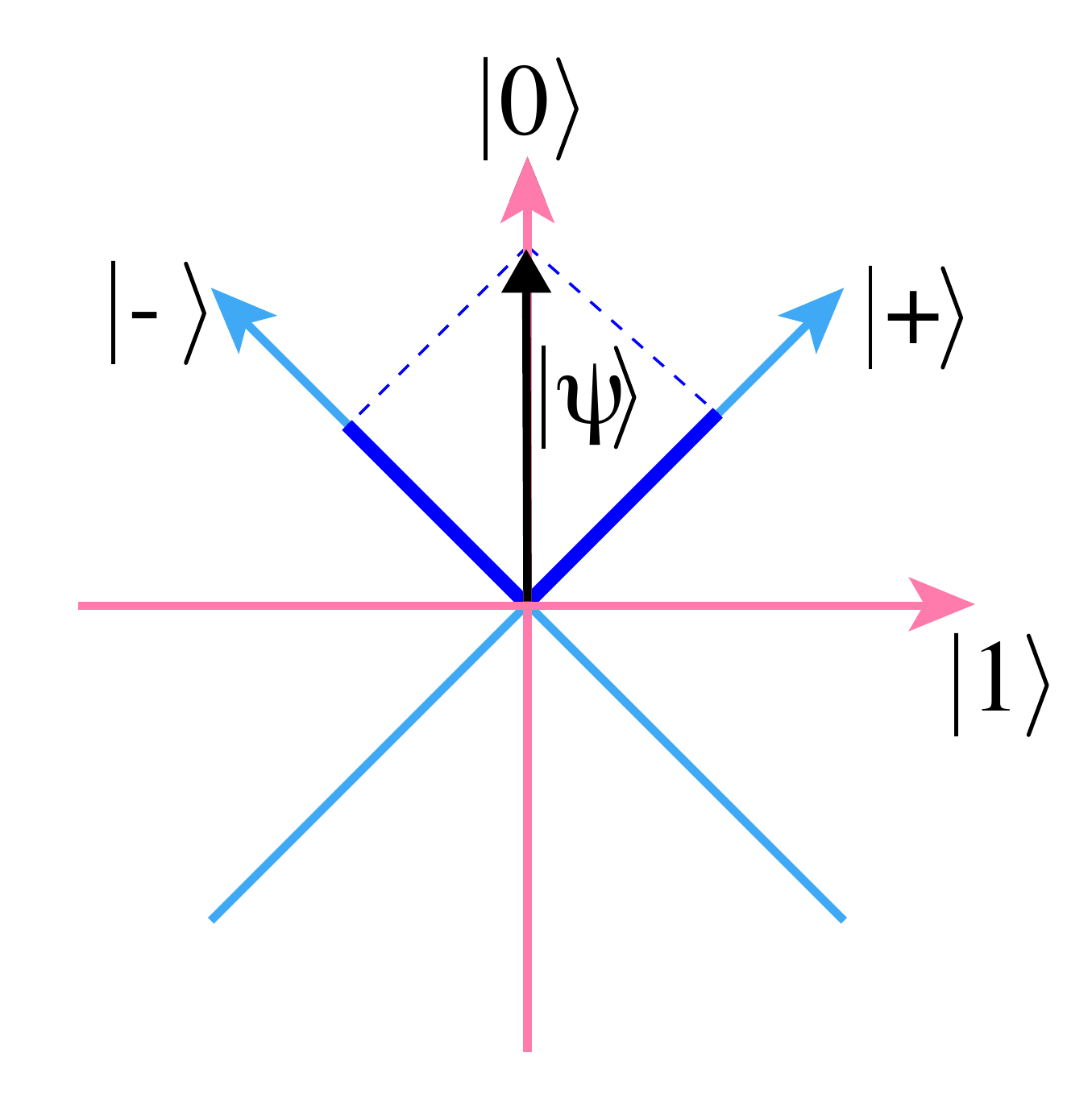}}
\caption{\label{fig:proiezioni} Two orthogonal references in the plane, to represent two different projective measurements: the computational basis $\{\ket0, \ket1\}$, and the  basis $\{\ket+, \ket-\}$ defined in \eqref{eq:diag-basis}. (a) A generic vector, with its components on the two references. (b) A special case of an eigenvector of the first reference which has equal components on the second reference.}
\end{figure}

If we consider the Hilbert space representing the states of the quantum system, each basis can be seen as a different POVM. In particular, an orthogonal basis will correspond to a projective measurement. 
The preparation and measurement of the quantum state of a physical system can be described in the language of QIT in terms of the encoding and decoding of information by two parties, traditionally called Alice and Bob.
The quantum complementarity is then related to the choice of the basis in which each party operates. If we consider the example of a qubit, in figure \ref{fig:proiezioni} two different orthogonal bases are shown, the  \emph{computational basis} $\{\ket0, \ket1\}$, and the  basis $\{\ket+, \ket-\}$, where
\begin{subequations} \label{eq:diag-basis}
\begin{align}
\ket+ &= \frac1{\sqrt2} (\ket0+\ket1)\\
\ket- &= \frac1{\sqrt2} (\ket0-\ket1).
\end{align}
\end{subequations}

Alice may choose to encode some information in the qubit, using the computational basis $\{\ket0, \ket1\}$, i.e. she prepares the system in one of the two states of this basis (see figures \ref{fig:qubit} and \ref{fig:proiezioni}). The  qubit will be then transmitted to Bob, who will perform a measurement to decode the information. If he chooses the diagonal basis $\{\ket+, \ket-\}$, he will be in the situation where both outcomes of the measurement have 0.5 probability (see figure \ref{fig:proiez2}). To describe this situation in  terms of information we can use the concept of mutual information expressed in  \eqref{eq:mutu-info}, and say that the mutual information between the (classical) random variable representing Bob's measurement outcome and the (classical) random variable representing the information encoded by Alce, is zero. This means in other terms that the Bob can not access the information of Alice.
This situation  expresses the concept of quantum complementarity, and based on this concept Charles Bennett and Gilles Brassard in 1984 devised the idea of \emph{quantum cryptography} \cite{Bennett-84}, which over the years has become one of the most developed applications of QIT \cite{Bennett-92a, Grosshans-03, SARG-04, Grazioso-13b, Ushenko-15}.

\subsection{von Neumann entropy}
In analogy to the definition of \emph{information content} of a classical random variable (Shannon entropy) defined in \eqref{eq:Sh-entropy}, it is possible to define the von Newmann entropy, in the case of a quantum random variable, in the following way:
\begin{equation} \label{eq:Vn-entropy}
S(\hat{\rho}) = \frac1{\log 2}\ Tr\left( \hat{\rho} \log \frac{1} {\hat{\rho}} \right).
\end{equation}
Here $Tr(\cdot)$ represents the trace of the density matrices, and $\rho$ is the density operator representing the random variabile of which $S$ represents the (quantum) information content.

\subsubsection{Quantum evolution}
\label{sec:evolution}
To complete the axiomatic framework of quantum mechanics we need one last postulate, about the evolution of a quantum system. It states that the evolution in time of a quantum system is described by an unitary transformation over the Hilbert space describing the states:
\begin{equation}
\ket{\psi(t)} = \hat{U} \ket{\psi(0)}.
\end{equation}
Here we will not  give  the details about the actual unitary operator, described by Shr\"odinger equation.

\subsection{Holevo theorem (Holevo bound)}

One of the most important results of QIT is the following theorem, called after Alexander Holevo \cite{Holevo-73}. As for the description of quantum complementarity, this result is best described in terms of the interaction between the two parties Alice and Bob. 
\begin{theorem}[Holevo bound]
Let's suppose that Alice prepares the quantum system in a mixed state described by the density operator $\hat{\rho}_{X}$, where $X = \{\ket{x_{1}}, \ldots, \ket{x_{n}} \}$ are the possible pure states, and $\{ p_{1}, \ldots, p_{n} \}$ are the corresponding probabilities. Then, Bob performs a measurement, described by a POVM built (as described in section \ref{sec:POVM}) on the set of pure states $Y = \{\ket{y_{1}}, \ldots, \ket{y_{n}} \}$, and we denote $y$ the outcome of this measurement.
It is possible to prove that for any such measurements Bob may do there is an upper bound for the \emph{mutual information}  \eqref{eq:mutu-info} between the two random variables $X$ and $Y$. In particular:
\begin{equation} \label{eq:Holevo}
I(X:Y) \leq S(\hat{\rho}) - \sum_{x} p_{x} S(\hat{\rho}_{x})
\end{equation}
where $\hat{\rho} = \sum_{x} p_{x} \hat{\rho}_{x}$ is the density operator describing the global mixed state prepared by Alice.
\end{theorem}

It is worth to stress that from the point of view of Alice (the sender), the information she encodes in the system is a classical information. We can represent it as the integer index labelling the states in the  set of quantum states $X = \{\ket{x_{1}}, \ldots, \ket{x_{n}} \}$ chosen for the encoding. On the other hand, from the point of view of  Bob(the receiver), the system is in a quantum mixed state. The following theorem expresses the relationship between the information contained in those two random variables.
\begin{theorem}
Given a classical random variable, encoded in a quantum system using the set of pure states $X = \{\ket{x_{1}}, \ldots, \ket{x_{n}} \}$, the relation between the information contained in this classical random variable, and the quantum information contained a   mixed quantum state $\hat{\rho}_{X}$ built with those pure states is:
\begin{equation}
S(\hat{\rho}) - \sum_{x} p_{x} S(\hat{\rho}_{x}) \leq H(X)
\end{equation}
the equality being reached in the case $\{\ket{x_{1}}, \ldots, \ket{x_{n}} \}$ are all orthogonal vectors. 
\end{theorem}
Because of this second  result, we can express the Holevo theorem \eqref{eq:Holevo} saying that in a quantum encoding-decoding process the amount of information that Bob can access is in general less than the (classical) information initially encoded by Alice, and that this information can be fully accessed only in the special case where the  set of quantum states used for the encoding is orthogonal.

\subsection{No-cloning theorem}
\label{sec:no-cloning}
Another important result of QIT is the no-cloning theorem, introduced by and Wootters, Zurek and  Dieks in 1982 \cite{Wootters-82, Dieks-82}. It is a no-go theorem that can be stated very briefly as follows: 
\begin{theorem}[No-cloning]
It is impossible to create an identical copy of an arbitrary \emph{unknown} quantum state.
\end{theorem}

The crucial part is the fact that the theorem applies to a situation where the state is unknown.

The theorem can be expressed also in the following alternative statement:
\begin{theorem}[No-cloning]
Given two  states $\{\ket{\psi_{1}}, \ket{\psi_{2}}\} \in \mathcal{H}$, which are non-orthogonal, i.e. $0 < |\braket{\psi_{1}}{\psi_{2}}| < 1$, it doesn't exist an unitary transformation defined on two states $\hat{U}: \mathcal{H}\otimes\mathcal{H}\rightarrow\mathcal{H}\otimes\mathcal{H}$ such that 
\begin{equation}
\hat{U}(\ket{\psi_{i}} \ket{0}) = \ket{\psi_{i}} \ket{\psi_{i}} 
\end{equation}
when $i$ is not  known, i.e. when $\psi_i \in \{\psi_1, \psi_2\}$ is unknown.
\end{theorem}

\section{The Black Hole Information Paradox}

\subsection{Black holes}

For the purpose of this review, black holes (BHs) can be briefly described as  objects so dense, and with a gravitational field so strong, that on a surface external to them, and called \emph{events horizon}, the escape velocity is higher than the speed of light. This implies that no physical object, not even light itself, can ever leave a BH once it is inside its event horizon.

\subsection{Hawking radiation and black hole evaporation}

\begin{figure}[!htbp]
	\begin{center}
	\includegraphics[width= \columnwidth]{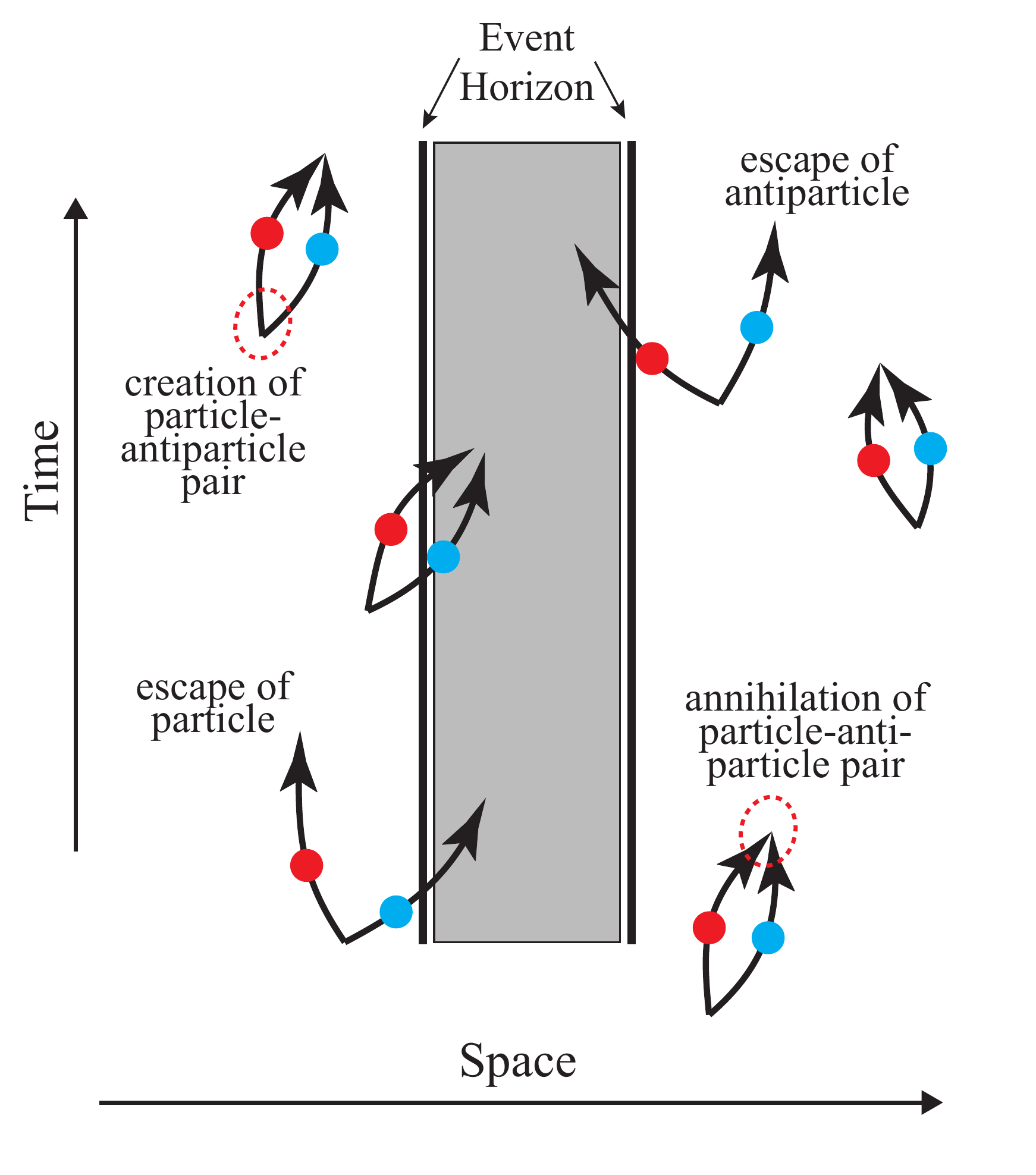}
	\caption{\label{fig-HR} Schematics of the mechanism of quantum vacuum fluctuation and generation of Hawking radiation.}
	\end{center}
\end{figure}

The work of Stephen Hawking in 1974 \cite{Hawking-74} introduced the notion of the \emph{Hawking radiation} (HR). This phenomenon is in turn due to the phenomenon of \emph{quantum vacuum fluctuations}, that was discussed and theorized at the beginning of the 20th century by the scientists that contributed to develop quantum theory (see e.g.\ \cite{Debye-13, Nernst-16}).   Quantum vacuum fluctuations are in turn linked to what has been subsequently formalized as the \emph{Heisenberg uncertainty principle} \cite{Heisenberg-27, Cohen1}, and can be summarized as the continuous and very rapid creation and annihilation of particle-antiparticle  couples  (see  figure \ref{fig-HR}). 
Hawking theorized that there is a non-zero probability that a  particle-antiparticle couple is generated close enough to the BH's event horizon, so that one of the two particles manages to escape before they re-annihilate while the other is trapped inside the horizon. The net effect is  a radiation emitted from the BH while taking some energy from it, and because of the mass-energy equivalence, the phenomenon can be described as the evaporation of the BH.
The Hawking radiation has an extremely low intensity, but if the BH is small enough, it can lead to the complete evaporation of the BH in a physically meaningful time, compared to the age of the universe. 
In its subsequent detailed quanto-mechanic calculations \cite{Hawking-75, Hawking-76b}, Hawking showed also  that the quantum state in which the HR is emitted is a \emph{maximally mixed state} (see section \ref{sec:density-mixed}). 

\subsection{Black hole paradox}

Since it is always possible to prepare the BH, as soon as it forms, in a pure state, and then leave it isolated, the phenomenon of HR leads to a contradiction.
Indeed if we  consider an  isolated BH as an isolated quantum system,  according to the postulates of QM seen in section \ref{sec:evolution}, its evolution should be described by an unitary transformation. But if we consider the process of complete evaporation of the BH, and take into account that the HR is emitted in a mixed state, we would have the evolution of an isolated quantum system from a pure state to a mixed state, in contradiction with that postulate. For what follows it is worth to remember that a maximally mixed state is such that each state of the mixture is equiprobable. So if we describe the final state of the Hawking radiation after the complete evaporation as a quantum random variable, this is in a maximally mixed state, and therefore it has \emph{zero mutual information} with the quantum random variable describing the initial state.

\subsubsection{BH paradox in terms of QIT}

It is possible to rephrase this contradiction  using the concepts of quantum information theory, so to show that contradicting the postulate of unitary evolution of an isolated quantum system is equivalent to contradict the no-cloning theorem introduced in section \ref{sec:no-cloning}.

\begin{figure}[!htbp]
	\begin{center}
	\includegraphics[width=\columnwidth]{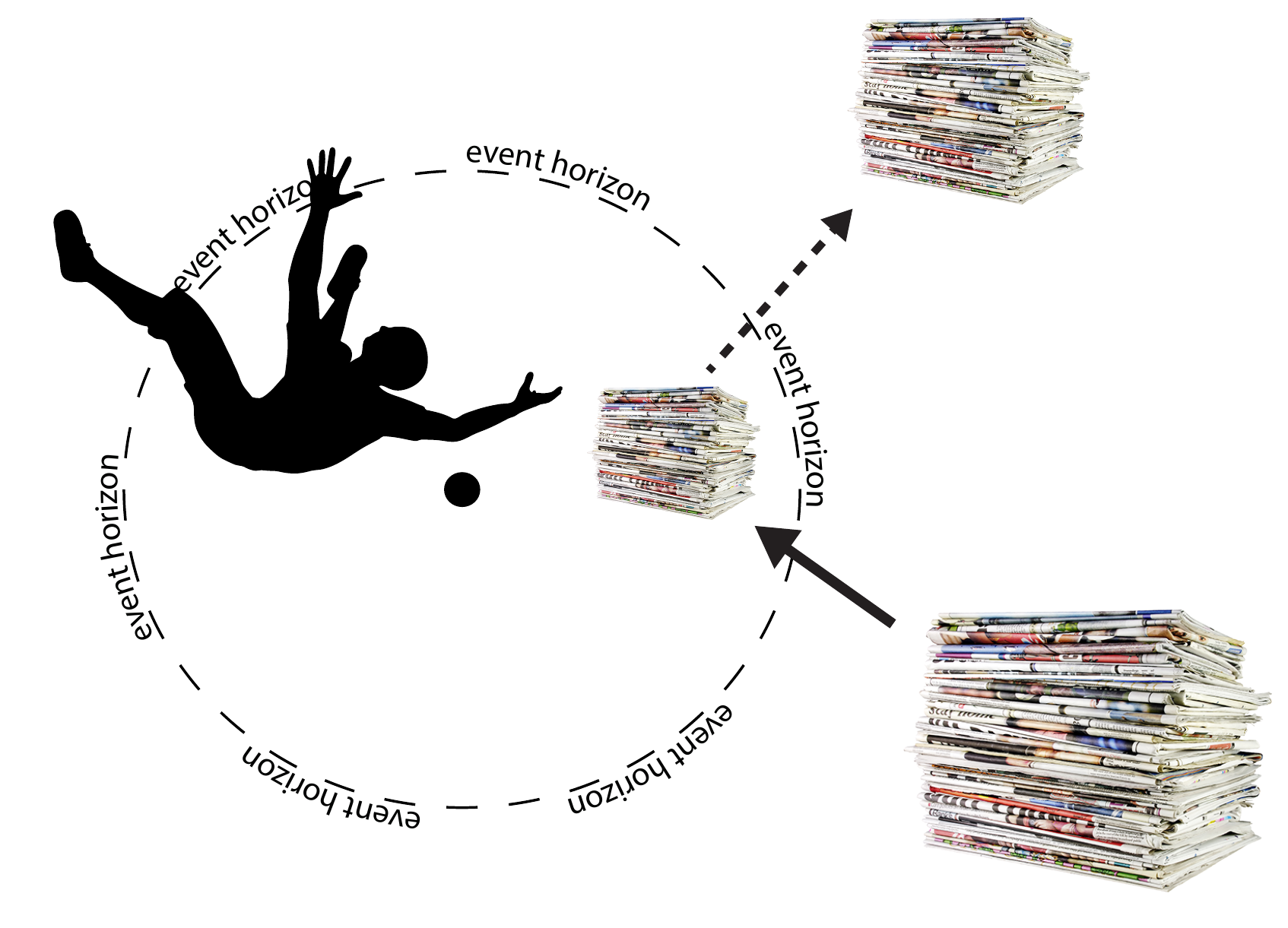}
	\caption{\label{fig:BH} Information falling into the event horizon: can it, even in principle, be retrieved? From the point of view of an in-falling observer, crossing the event horizon has no physical effect, and this suggests that also the information is not destroyed when it falls inside the horizon.}
	\end{center}
\end{figure}

Let's consider a physical system, containing a certain amount of  \emph{information},  dropped into the BH at an early time, and let's ask the question whether this information can in principle be retrieved  at a later time or not (see  figure \ref{fig:BH}). 

In a deterministic system, following the dynamic equations that describe its evolution, it is in principle possible to reconstruct an earlier state once we fully know the state at a later time (with emphasis on the \emph{full knowledge} of any degree of freedom and their correlations). So, if a BH is well described by quantum mechanics, the answer to the question about the information retrieval should be affirmative, and the Hawking radiation is a good candidate to explain how the information can escape. This in turn  would question general relativity, from which the very definition of event horizon descends \cite{Misner-73, Birrell-84}, because by definition nothing can escape the event horizon.\\
If on the other hand the answer to the question about the information retrieval is negative, then it means that the quantum-mechanical description of the BH and its evolution has to be revised.

Moreover, we can show how, if we accept the notion that somehow the information initially dropped inside the event horizon, eventually  escapes via the Hawking radiation, we incur in another problem.
Indeed, from the point of view of an in-falling observer, crossing the event horizon has no physical effect. So we can safely assume that the information dropped in the BH still exists intact, inside the event orizon (at least until it reaches the internal singularity of the BH).

Therefore, if the information also escapes, it means that  at least a finite time, two copies of the same information exist, inside and outside the event horizon. So this would contradict the no-cloning theorem of section \ref{sec:no-cloning}.

\subsubsection{Contributions to the solution from QIT}

Although the BH information paradox is still an open problem, QIT has contributed to its comprehension with some important results and insights.

Jacob Bekenstein is one of the leading authors of such line of research \cite{Bekenstein-04}. In 1972 he has introduced a \emph{generalized second law}  describing the thermodynamics of BHs \cite{Bekenstein-72}, and in the 1973 he has introduced a definition of BH's \emph{entropy}, as being proportional to its area $\mathcal{A}$ and inversely proportional to the square of Plank's length $\ell_{P}^{2}$:
\begin{equation}
S_{BH} \propto \frac{\mathcal{A}}{\ell_{P}^{2}}.
\end{equation}
Then, at first Bekenstein \cite{Bekenstein-81}, and then Bousso \cite{Bousso-02} have found upper bounds for the BH's entropy.
Since the double meaning of the entropy as both a thermodynamic parameter and a measure of the information content of a system (see section \ref{sec:info}) these results have suggested a  information theoretical approach to solve the paradox.

Hayden and Preskyll \cite{Hayden-07} have used results from quantum error correction, to extend a result already found by Page \cite{Page-93b}. When the BH is in an advanced stage of its evaporation, more precisely when its entropy is less than  half the initial amount, they prove that the information retention time, i.e. the time needed for the information dropped in the event horizon to re-emerge in the Hawking radiation, is relatively short, and in particular: 
\begin{equation}
t_{\text{info}} = \mathcal{O}(r_{S} \log r_{S} )
\end{equation}
where $r_{S}$ is the Schwarzschild radius.\\

Another contribution to the solution of the BH information paradox, also used by Hayden and Preskyll, is the concept of BH complementarity \cite{Susskind-93, tHooft-85}. This approach considers  two possibilities:  the information traveling toward the BH from outside, when reaches the event horizon is  either transmitted inside or reflected outside. Then, the suggestion is that instead of choosing between those two possibilities, we can accept them both. To solve the conflict with the no-cloning theorem, we assume that, because of the \emph{quantum complementarity} discussed in section \ref{sec:complementarity} it is impossible for any observer to observe both descriptions, or access both copies of the information. An external observer  will see the incoming information being absorbed by the event horizon, and then re-transmitted outside by means of the Hawking radiation, all this process being unitary. The observer falling inside the event horizon from outside will not notice the crossing,  and will continue to observe the information that is falling with him. But he will not be able to access the information reflected outside with the Hawking radiation, because that will be encoded in a different basis, such that the mutual information is zero.

Another important result worth to mention is the \emph{holographic principle}, a general result which can be stated as follows: \emph{``Physical processes in a system of $\mathcal{D}$  dimensions are reflected in processes taking place on the $\mathcal{D}-1$ dimensional boundary of that system. There is an equivalence between theories of different sorts written in space-times of different dimensions''} \cite{tHooft-00, Bekenstein-04}.

The fields of QIT, Astrophysics and general relativity have all gained from this interdisciplinary approach; as an example the concept of Generalized Second Law, and the Holographic Principle have also  lead to results in QIT. In particular, upper bounds have been proven for the entropy outflow $\frac{\partial S}{\partial t}$, which is a proxy for the communication rate, or information channel capacity \cite{Bekenstein-04}.

\section{The renormalization group information flow}

\subsection{Description of the RG}

The main idea of the renormalization group (RG) is that of a tool to extract the macroscopic description of a physical system (e.g.\ a field) from its microscopic model. First of all, the change in the descriptions going from the microscopic to the macroscopic model is captured by the change of the  interaction constant $g(\mu)$ in the interaction term of the hamiltonian. 

This change can be described as the action of an operator $\hat{G}$  applied to the interaction constant:
\begin{equation} \label{eq:RGequation}
g(\mu_{2}) = \hat{G}\left[ \ g(\mu_{1}) \right] 
\end{equation}
where  $\mu_{i}$ is a parameter that represents the different scales. Although this transformation is called ``renormalization group'', it is not formally a group. It is just a "flow of transformations" in the space of all the possible hamiltonians. The main reason why the RG is not a group, is that given a transformation from a small scale description to a large scale description, the inverse transformation is not necessarily defined.

In 1954 Murray Gell-Mann and Francis Low published a work on quantum electrodynamics (QED) \cite{Gell-Mann-54}, in which they studied the photon propagator at high energies.
They introduced the concept of scaling transformation with a group-like formalism, where the group operator $\hat{G}$ transforms the electromagnetic coupling parameter $g$:  
\begin{subequations} \label{eq:Gell-Mann-Low}
\begin{align} 
& \hat{G}\left[g(\mu_{2})\right] = \left(\frac{\mu_{2}}{\mu_{1}}\right)^{d}\ \hat{G}\left[g(\mu_{1})\right] \\
& g(\mu_{2}) = \hat{G}^{-1}\left[\left(\frac{\mu_{2}}{\mu_{1}}\right)^{d}\ \hat{G}\left[g(\mu_{1})\right] \right] .
\end{align}
\end{subequations}
Equation \eqref{eq:Gell-Mann-Low} expresses the requirement that before and after the scaling, the physical laws don't change. So the equation requires that the coupling parameter before and after the scaling changes taking into account the scaling factor $\left(\frac{\mu_2}{\mu_1}\right)^d$. Going from this discrete scaling $\mu_{1} \rightarrow \mu_{2}$ to a continuous scaling transformation, it is possible to define  a function $\beta(g)$ that expresses the corresponding continuous transformation of the coupling parameter $g$:
\begin{equation} \label{eq:beta}
\beta\left[g(\mu)\right] = \frac{\partial g(\mu)}{\partial \ln(\mu)}.
\end{equation} 
Between 1974 and 1975 Kenneth Wilson and John Kogut introduced a more general description of this idea \cite{Wilson-74, Wilson-75a, Wilson-75b}.  In this description, the large scale (macroscopic) behaviour will be linked to the low energy regime of the model, because at long distance  only  long wavelengths are relevant, while for the microscopic behaviour higher energies will be relevant. With reference to this, in the language of the RG the microscopic, high energy model will be called the \emph{ultraviolet limit}, while the macroscopic, low energy one will be called the \emph{infrared limit}. Another language to express the description at different scales is in terms of \emph{fine graining} and \emph{coarse graining}.

To give an example of the low energy approximation, we can imagine a  sinusoidal potential for the microscopic model, and its approximation with  a parabolic potential for the macroscopic description. This will be  a good description at low energies, i.e.\   at the bottom of the microscopic potential. However, at high energies this approximation may introduce some divergencies, involving as an example the integration over bigger ranges of energies.  Since those divergencies are only due to the approximated description of the potential, this can be corrected introducing a cut-off for the high range of energies.
\begin{figure}[!htbp]
	\subfloat[\label{fig:renorm-abstr}]
		{\includegraphics[width=0.5\columnwidth]{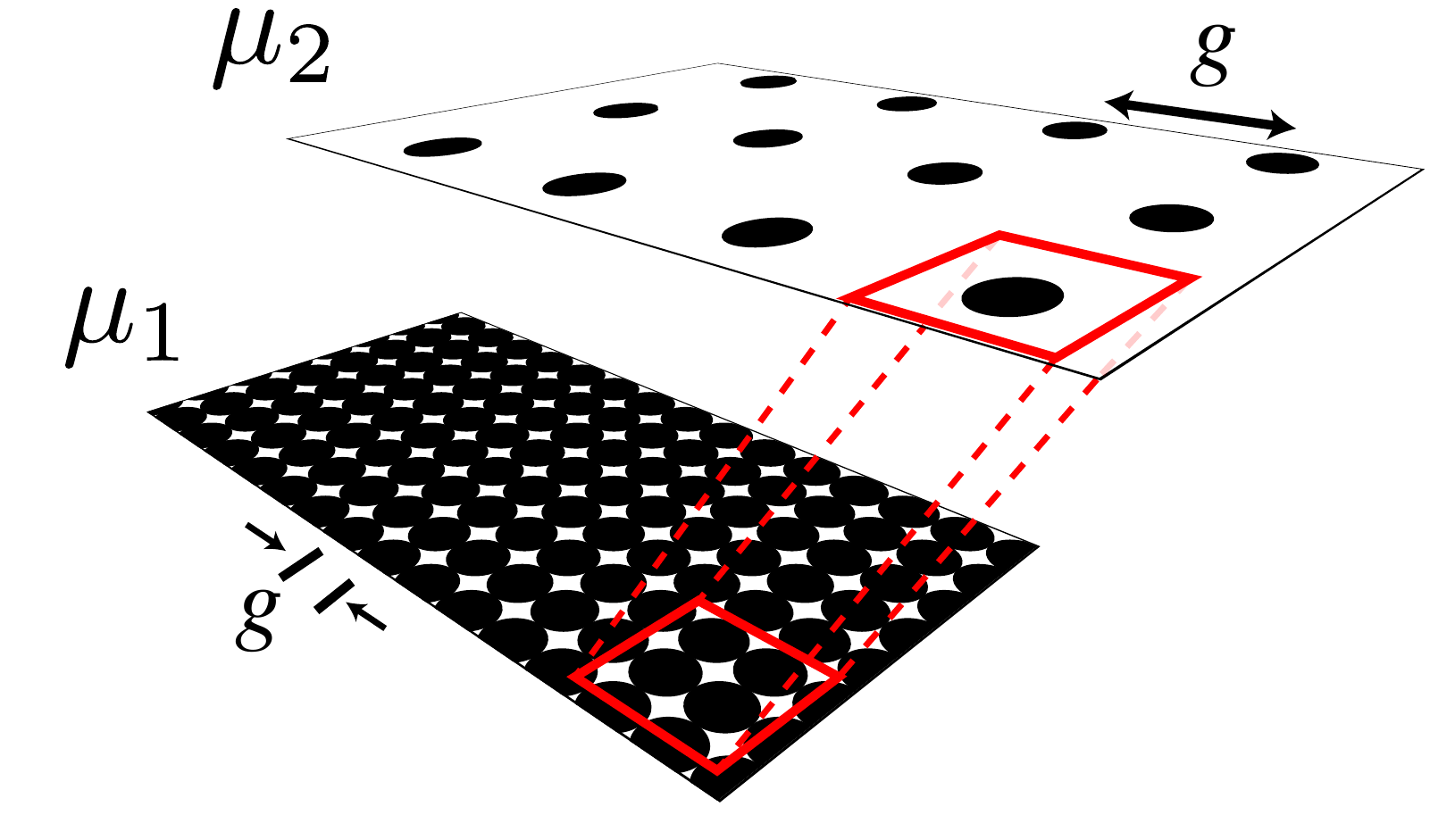}}
	\subfloat[\label{fig:renorm-astro}]
		{\includegraphics[width=0.5\columnwidth]{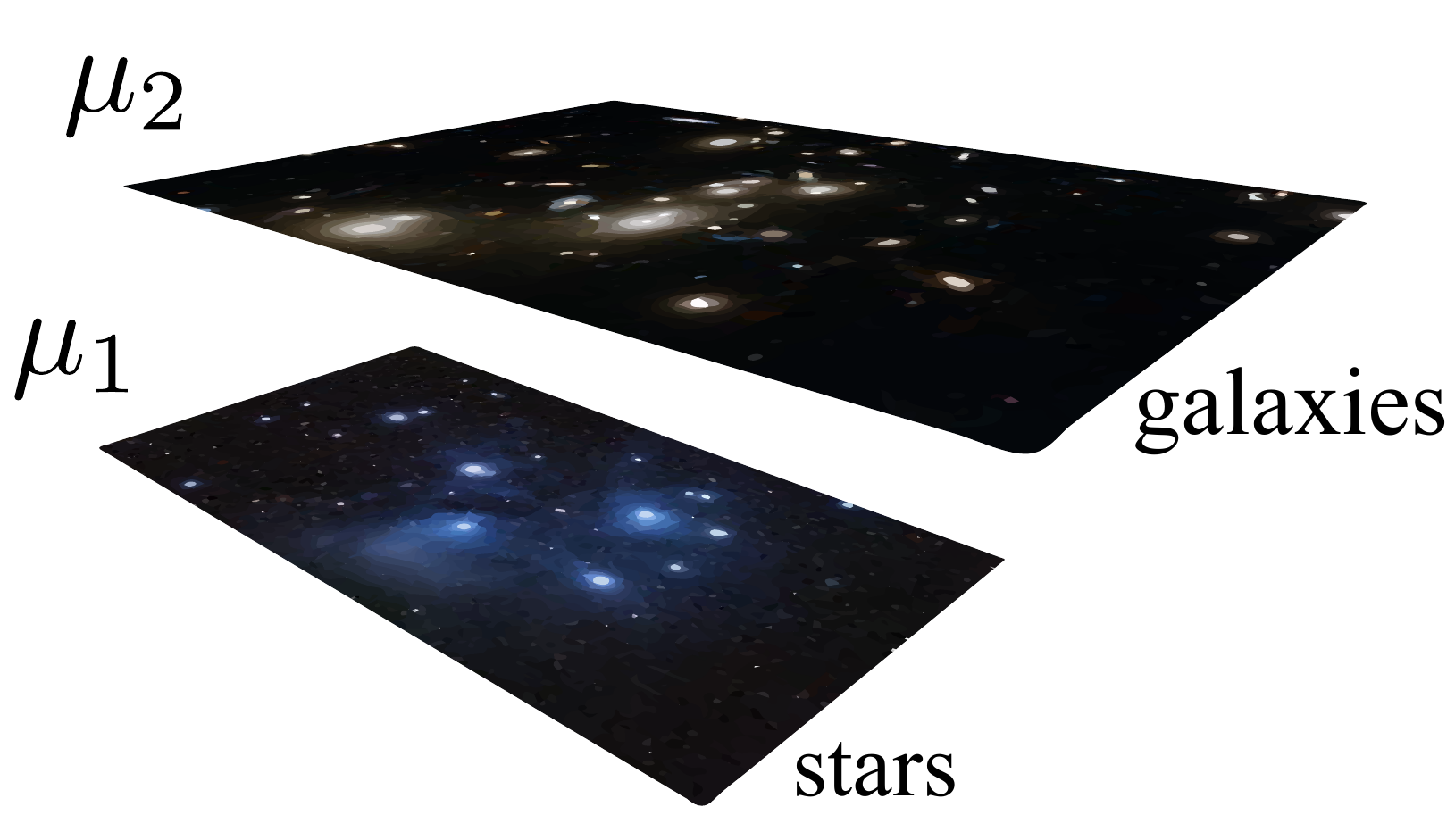}}
	\caption{\label{fig:renormalization} abstract description of the renormalization group. (a) Two different scales of modelling, with two different interacting constants. (b) An example of such different scales can be found in astrophysics, where the description at the scale of stars (lower image) has an interaction constant different from the description at the scale of galaxies (upper image).}
\end{figure}
The dynamics of a composite system can be described by the interactions between its components.  At a certain scale (graining) $\mu_{1}$  the physics of that model is described by the hamiltonian of the system, and in particular by its  interaction term, i.e.\ by the interaction constant $g(\mu_{1})$.
At an bigger scale (coarse graining) $\mu_{2}$, the  components of the lower scale can be ``clustered'' into a single element of the coarse graining (see section \ref{fig:renorm-abstr}), and the  interaction constant is in principle changed. The equations  expressing the constrain that: ``the physics at different scales has to be the same'' are \eqref{eq:RGequation} and \eqref{eq:Gell-Mann-Low}, which express the constrains for the interaction constant $g(\mu_{i})$, and another equation that express the constrain between the correlation at different scales, which is the  the Callan-Symanzik equation \cite{Callan-70, Symanzik-70, Symanzik-71}:
\begin{equation}
\left[ m \frac{\partial}{\partial m} + \beta(g) \frac{\partial}{\partial g} + n \gamma\right] C^{(n)} (x_1, \ldots, x_n ; m, g) = 0
\end{equation}
where: $m$  is the mass, $C$  is the correlation function between the $(x_{1}, \ldots, x_{n})$ elements of the system, $\beta$ and $\gamma$  are two functions that ``compensate'' the effect of the scale change, in order for the  description (i.e.\ the correlation function) at the different scales to be consistent. In particular $\beta$, which we have already seen in \eqref{eq:beta}, captures the change of the coupling constant, while $\gamma$ captures the change of the field itself.

In applying the group transformations, we go from one point of the space (manifold) of all the possible hamiltonians (i.e.\ in the manifold of the $\beta$s and $\gamma$s) to another. However, there are some points, called \emph{critical points}, or \emph{conformal points}, where the function $g(\mu)$ has its minimum. From another point of view we can think at the manifold  of the hamiltonians (each describing a different model for the system, at different scales, with different values of the coupling constant), and then think that the RG transformations describes a flow from one model to the other. The flow always ends at the points that are invariant for this transformation, so those points have to be self-similar. Each of the critical points    are characterized by the (minimal) value that the function assumes there, and this value is called the "central charge" of the system. 

\subsection{The c-function and the link to QIT}

The \emph{c-theorem} of Alexander Zamolodchikov \cite{Zamolodchikov-86} individuates, in the case of a two-dimensional renormalizable field, a function which is monotonic  along the RG transformations.

This monotonicity suggests an information theoretical meaning for this function, analogue to the information content.  \cite{Apenko-12, Beny-12, Beny-13}.

Since the seminal result by Zamolodchikov, several authors have worked on c-theorems at dimensions higher than 2 \cite{Cardy-88, Jack-90, Cappelli-91, Appelquist-99}.

Another approach to the RG is the density matrix renormalization group (DMRG) \cite{White-92, Noack-99}. 
Osborne and  Nielsen \cite{Osborne-02} make more explicit the link between  DMRG and QIT.
A characteristic feature of  critical phenomena is the emergence of collective behaviour, and it is conjectured that  quantum entanglement is the origin of this cooperative behaviour. DMRG and its explicit quanto-mechanical approach seems the ideal formalism with which to substantiate this conjecture \cite{Aharonov-00, Sachdev-11}.

Finally, a different interdisciplinary approach, not necessarily linked to information theory, is the parallel between  the renormalization used in quantum field theory and the renormalization used in thermodynamics and statistical mechanics to describe critical phenomena \cite{Wilson-74, Parisi-98}.

\section*{Acknowledgements}
The author thanks Frédéric Grosshans, Gilles Brassard and Patrick Hayden for the fruitful discussions and inputs on  information theory,  the latter in particular for having introduced him to the black hole information paradox, and Spyros Sotiriadis for the discussions on the renormalization group.

\bibliography{biblio-QIT_paper}

\end{document}